\documentclass[12pt]{article}

\usepackage[textheight=240mm,textwidth=169.69mm]{geometry}
\usepackage{graphicx} % Required for inserting images
\usepackage{natbib}
\usepackage{amsmath,amssymb,amsfonts,amsthm}
\usepackage{xcolor}
\usepackage{multirow}
\usepackage{makecell}
\usepackage{siunitx}
\usepackage{booktabs}
\usepackage{hyperref}
\hypersetup{
    colorlinks=true,
    linkcolor=blue,
    filecolor=blue,      
    urlcolor=blue,
    citecolor=blue
    }
\usepackage{anyfontsize}

\usepackage{newtxtext}
\usepackage{newtxmath}

\title{Surface bubble lifetime in the presence of a turbulent air flow, and the effect of surface layer renewal}
\author{Tristan Aur\'egan$^1$ and Luc Deike$^{1,2}$}

\date{{\small $^1$ Department of Mechanical and Aerospace Engineering, Princeton University, Princeton, New Jersey, USA\\
$^2$High Meadows Environmental Institute, Princeton University, Princeton, New Jersey, USA}}

\begin{document}

\maketitle

\begin{abstract}
    Surface bubbles in the ocean are critical in moderating several fluxes between the atmosphere and the ocean. In this paper, we experimentally investigate the drainage and lifetime of surface bubbles in solutions containing surfactants and salts, subjected to turbulence in the air surrounding them modelling the wind above the ocean. We carefully construct a setup allowing us to repeatably measure the mean lifetime of a series of surface bubbles, while varying the solution and the wind speed or humidity of the air. To that end, we show that renewing the surface layer is critical to avoid a change of the physical properties of the interface. We show that the drainage of the bubbles is well modelled by taking into account the outwards viscous flow and convective evaporation. The mean lifetime of surface bubbles in solutions containing no salt is controlled by evaporation and independent on surfactant concentration. When salt is added, the same scaling is valid only at high surfactant concentrations. At low concentrations, the lifetime is always smaller and independent of wind speed, owing to the presence of impurities triggering a thick bursting event. When the mean lifetime is controlled by evaporation, the probability density of lifetime is very narrow around its mean, while when impurities are present, a broad distribution is observed.
\end{abstract}

\section{Introduction}

Bubbles at the surface of the ocean play a key role in atmosphere-ocean interactions. When they burst, they release droplets into the air that facilitate heat and mass exchanges \citep{deike_mass_2022}. These droplets carry with them the chemical and biological contents of the top layer of the ocean: salts but also any organic or artificial contaminants that may be present \citep{cunliffe_sea_2013,burrows_physically_2014,veron_ocean_2015}. The smallest droplets may be entrained in upper layers of the atmosphere where they serve as an important source of cloud condensation nuclei \citep{lewis_sea_2004}. Understanding and being able to predict the number and size of these droplets is therefore important to better model atmospheric processes. 

There are several pathways for the generation of droplets from bubble bursting \citep{lewis_sea_2004,jiang_submicron_2022,deike_mechanistic_2022,jiang_abyss_2024}, but the one we are most interested in is the ``film drops'' mechanism as it may be able to produce very small droplets associated with a large uncertainty \citep{dubitsky_effects_2023,zinke_effect_2022}. Film drops are generated during the retraction of the bubble cap, through a centrifugal instability \citep{blanchard_film_1988,lhuissier_bursting_2012,jiang_submicron_2022} and are typically of similar size as the thickness of the cap from which they were generated. It is therefore important to know the drainage rate and lifetime of those surface bubbles in order to estimate the size of droplets generated. The drainage of surface bubbles has been thoroughly investigated by many laboratory studies in various conditions \citep{burger_persistence_1983,struthwolf_residence_1984,lhuissier_bursting_2012,champougny_life_2016,poulain_biosurfactants_2018,poulain_ageing_2018,miguet_stability_2020,pasquet_impact_2022,shaw_film_2024}. In the conditions relevant to oceanic surface bubbles, the drainage is driven by a curvature induced pressure gradient, resulting in the thickness of the bubble cap $h$ decreasing like $t^{-2/3}$, until at long times (a few tens of seconds) direct evaporation from the cap dominates, and the thickness decreases linearly in time. Corrections can be added to this base flow to account for surface tension gradients that can form because of evaporation from the top of the bubble \citep{chandransuja_evaporationinduced_2018,poulain_ageing_2018,shaw_film_2024}.

The lifetime of surface bubbles ranges between a few to tens of seconds, but precisely predicting the expected lifetime for a given set of physical parameters is challenging. Previous authors have noted the difficulty in even measuring lifetime as it is extremely sensitive to the type and amount of surfactants present in the solution and can vary drastically without any change of the physical parameters of the study \citep{poulain_ageing_2018}. The maximal lifetime that a bubble can live is set by evaporation as the fast removal of fluid from the cap at late times make it highly susceptible to bursting \citep{miguet_stability_2020,roux_everlasting_2022}. The difficulty in accurately predicting lifetime lies in the identification of a single bursting criterion: previous studies have shown that bubbles can burst at a wide range of thicknesses, some much larger than the thickness at which Van der Waals forces act. One possible explanation for bursting events at large thickness is proposed in \citet{neel_spontaneous_2018} or \citet{poulain_ageing_2018} as the formation of a hole in the cap due to sufficiently large local surface tension gradients caused by surfactants or impurities at the interface.

This study aims towards measuring mean lifetimes and drainage curves of surface bubbles in conditions relevant to the ocean. To this end, we use solutions representative of ocean water and introduce agitation in the air above the water surface. Indeed, the studies mentioned above investigated the behavior of bubbles in laboratory settings, with the water surface flat and motionless, but bubbles at the surface of the ocean are constantly pushed by the motion of the ocean surface as well as subjected to turbulence in the air from the atmospheric boundary layer. This turbulence potentially having the effect of altering drainage through bubble cap entrainment or altering the evaporative rate of bubbles. In addition, we recently showed that mixing in the air around thin films have the potential to drastically affect lifetime and reduce the large fluctuations observed in previous studies \citep{auregan_drainage_2025}. 

The capacity of oceanic aerosols to serve as cloud condensation nuclei strongly depends on their chemical composition \citep{burrows_physically_2014}. As a result, the exact composition of the solution in ions, surfactants, but also organic material may affect the final coupling between atmosphere and ocean. In this study, we use model solutions that can readily be prepared in the lab: containing either only NaCl in seawater concentration or a mixture of salts representative of those that can be found in the ocean. We use Sodium Dodecyl Sulfate (SDS) as a surfactant as it has often been used in the atmospheric aerosols' literature to model the naturally occurring lipids in the ocean \citep{li_influence_1998,tuckermann_surface_2007,burrows_physically_2014}. Other compounds such as palmitic or stearic acid are though to better model the oceanic surface layer \citep{burrows_physically_2014}, but SDS has the advantage of being easily available and a wide body of literature on its physical properties \citep{fainerman_surface_2010}. Finally, our solutions contain no organic macromolecules (polysaccharides or proteins) that are sometimes used in atmospheric chemistry studies. 

This paper is organized as follows: in Sec. \ref{sec:matmet} we present the experimental setup and in Sec \ref{sec:overflow} the steps that have been taken to ensure the repeatability of the lifetime measurement with surface layer renewal. Then in Sec. \ref{sec:results} we present mean lifetime and drainage data for solutions containing only surfactants and surfactants and salts. In Sec. \ref{sec:distribution}, we analyze the shapes of the lifetime distributions obtained with various solutions. Finally, in the discussion (Sec. \ref{sec:discussion}) we analyze the differences between those cases and propose a rationalization. 

In the following, we make the distinction between different classes of surface active compounds. To avoid confusions, we define here different terms for these classes (similarly to \citet{mcnair_exogenousendogenous_2025}): surfactants refer to surface active molecules that are commonly used detergents and that we have purposefully added to the solution. Co-surfactants are similar molecules that can be found in trace amounts in commercially available surfactants powders. Finally, impurities are compounds that have an influence on surface properties that can come from various sources in the environment (dust, tap water organics, etc.) and that we have not purposefully added. Adding salts has both the effect of introducing ions in the solution that may influence the surface properties through electrostatic interactions and also adds impurities that are present in the powder that we use.

\section{Material and Methods}\label{sec:matmet}

\begin{figure}
    \centering
    \includegraphics[width=0.6\linewidth]{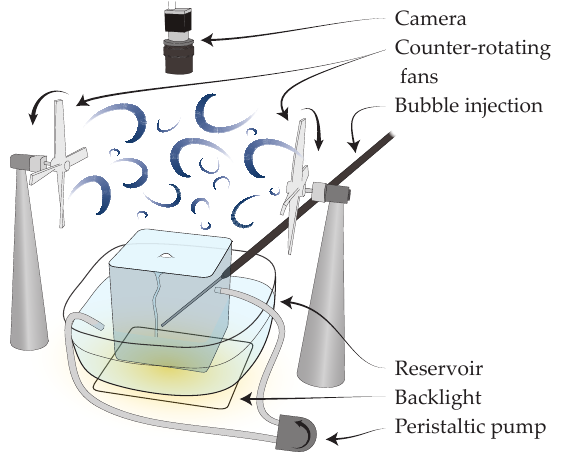}
    \caption{Schematic of the experimental setup. Bubbles are injected underwater through a needle connected to a syringe pump. Bubbles rise to the surface and are monitored via a top-down viewing camera and backlight system. Turbulence in the air is generated via two counter-rotating fans and the humidity of the air is also controlled. Finally, a peristaltic pump continuously overflows the container where bubbles are generated into a larger container, avoiding surface ageing.}
    \label{fig:setup}
\end{figure}

We measure the lifetime of surface bubbles in various conditions. The bubbles are injected 5 cm below the surface and rise freely up to the surface (70 mm $\times$ 70 mm) where they can move with no restrictions. The water level is slightly above the edge of the container such that the bubbles are naturally pushed away from the edges. The container is continuously overflowed into a larger reservoir using a peristaltic pump such that the surface layer is renewed constantly (detailed below). We record the bubbles with a camera (Basler acA1920-40um) placed above the surface, allowing us to measure the radius of the bubbles as well as their lifetime (from the moment they reach the surface to the final bursting event). A live detection algorithm detects the presence of a bubble on the surface from the camera feed and allows us to send a new bubble only when the previous one has burst. We run this experiment for a large number of bubbles, up until the relative uncertainty on the mean lifetime is below 5\% or at least 150 repetitions are reached.

The solutions used to generate bubbles are mixes of Sodium Dodecyl Sulfate (or SDS), with a surfactant concentration $c_\textsc{SDS}$ (the Critical Micelle Concentration or CMC of SDS is 8 mmol/L \citep{fainerman_surface_2010}) and salt (either pure NaCl or artificial sea salt (ASTM D 1141-98)). The conditions explored in this study are summed up in Tab. \ref{tab:solutions}. In addition to sodium and chloride ions, the sea salt solutions contain magnesium, calcium, sulfate and potassium ions in significant proportions. A small amount of the initial sea salt powder does not dissolve and remains as floating particles in the solution. SDS is a negatively charged surfactant and as a result, the combined effect on surface tension or other surface properties of a small amount of SDS and a small amount of ions in solution can be very different from each of the components alone. We measured the surface tension isotherms of all the solutions used in this study in a Langmuir trough using the Wilhemy plate method. These isotherms can be found in Appendix \ref{app:lang}. %Fig. \ref{fig:langmuir}.

The bubbles are generated by pushing air through a needle tip (inner diameter 1.6~mm) using a syringe pump, allowing us to obtain bubbles of cap radius $R = 2.5 \pm 0.2$~mm. Because of the large variations in surface tension, there is a slight trend towards smaller radii at low surface tension (except for the highest SDS concentration with sea salt where we decided to increase the syringe volume to obtain a radius closer to the mean value). %The effect of the slight radius changes are taken into account through the drainage timescale ($\tau_d$, defined in Eq. \eqref{eq:timescales}).

\begin{table}
    \centering
    \sisetup{
        table-column-width = 9mm,
        table-alignment-mode = none
    }
    \begin{tabular}{ c | *{4}{@{}S@{}} | *{5}{@{}S@{}} | *{6}{@{}S@{}}}
    \toprule
    Solution & \multicolumn{4}{c|}{No salt} & \multicolumn{5}{c|}{35 g/L NaCl} & \multicolumn{6}{c}{Sea salt} \\
    \midrule
    $c_{\textsc{SDS}}$ & 10 & 200 & 3000 & 10$^4$ & 10 & 50 & 100 & 200 & 400 & 10 & 20 & 50 & 100 & 200 & 400  \\
    $\gamma_0$ (mN/m) & 69.9 & 67.0 & 47.6 & 35.6 & 62.8 & 47.0 & 39.1 & 35.2 & 20.0 & 54.4 & 50.5 & 43.3 & 38.2 & 31.7 & 28.3  \\
    $R$ (mm) & 2.8 & 2.8 & 2.5 & 2.3 & 2.7 & 2.5 & 2.4 & 2.3 & 2.1 & 2.6 & 2.5 & 2.5 & 2.3 & 2.2 & 2.6  \\
    % $\tau_d$ (ms) & 48.6 & 54.7 & 59.4 & 79.7 & 45.9 & 61.6 & 68.7 & 79.4 & 157.9 & 61.0 & 59.7 & 71.5 & 75.9 & 78.3 & 219.6  \\
    \end{tabular}
    \caption{Physical parameters of the solutions used in this study: SDS concentration $c_\textsc{SDS}$, equilibrium surface tension $\gamma_0$ (measured using the Langmuir trough at zero compression), average cap radius of the bubbles generated $R$.} %, and associated drainage timescale $\tau_d$ (defined in Eq. \eqref{eq:timescales})}
    \label{tab:solutions}
\end{table}

The setup is placed in a container where the air is monitored. In particular, we want to monitor the relative humidity in the chamber, meaning the amount of water vapour in the air, relative to maximal amount the air could contain at a given temperature. The relative humidity ($\mathcal{R}_H$) defined as the ratio between the partial pressure of water vapour to the saturation pressure of water vapour, is therefore measured continuously using a dedicated sensor (HIH-4021-003). Its value is then duty-cycled controlled using a PID controller alternatively injecting dry or humid air in the chamber. This setup has been previously detailed in \citet{boulogne_cheap_2019}. We create a turbulent flow above the surface using two counter-rotating fans (5~cm radius) placed 17~cm apart on each side of the container (see Fig \ref{fig:setup}). The fans have four 3D printed blades and their angular velocity can be modulated. Throughout the study we use five conditions: 0, 50, 275, 460, and 610 rotations per minute. 

\begin{figure}
    \centering
    \includegraphics[width=0.65\linewidth]{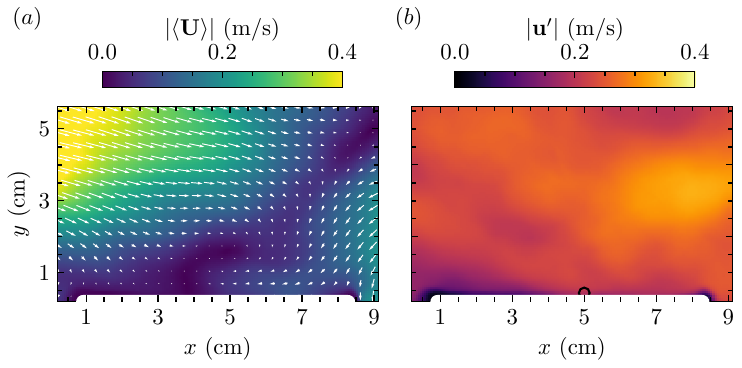}
    \caption{(a) Average flow velocity $\left< \mathbf{U} \right>$ above the bath surface. (b) Root mean square velocity $\mathbf{u}'$ above the bath surface. The withe area at the bottom is the location of the container, the black circle shows the typical size of bubble in this study ($R = 1.5$~mm). Both of these snapshots correspond to the case where the angular velocity of the fan is 460 RPM.}
    \label{fig:piv}
\end{figure}

The fans create a highly unsteady flow with a relatively low mean: for all conditions the root-mean-square velocity is larger than the mean velocity, especially close to the surface. We measured the properties of the flow generated using Particle Image Velocimetry. We seeded the flow with calcium carbonate particles of about 1 \textmu m and used a laser sheet in the vertical plane containing the centers of both fans to image the flow. The bath is in the middle between the two fans, resulting in a relatively weak mean flow $\left< \mathbf{U} \right>$ ($\left< \cdot \right>$ denotes a temporal average) with a slight downward component (Fig. \ref{fig:piv}, a). The magnitude of the root-mean-square (RMS) velocity $\mathbf{u}' = \left(\left<(\mathbf{U} -\left< \mathbf{U} \right>)^2 \right>\right)^{0.5}$ (Fig. \ref{fig:piv}, b) is homogeneous in space across the surface (along the $x$ direction). As we get close to the surface, the average flow decreases rapidly while the RMS value remains important. This means that the apparent flow over the bubbles has a very small mean and is essentially only fluctuating. As a consequence we believe that the large scale asymmetry in the mean flow visible in Fig. \ref{fig:piv} (a) does not influence the dynamics of the bubble.

% To quantify the flow over the surface, we measure relevant quantities for the convective evaporation by the turbulent flow. We compute the longitudinal structure function:
% \begin{equation}
%     D_{LL}(\mathbf{x}, \Delta r) = \frac14 \sum_{i = x,y} \sum_{j = \pm 1} \left< \left( \widehat{u_i}(\mathbf{x} + j \mathbf{e}_i \Delta r, t) - \widehat{u_i}(\mathbf{x}, t)\right)^2 \right>,
% \end{equation}
% with $\mathbf{\hat{u}} = \mathbf{U} -\left< \mathbf{U} \right>$ and $\mathbf{e}_i$ the unit vector in the $i$ direction (plotted in Fig. \ref{fig:piv} (c)). 
The integral length scale $L_{\rm int}$ is computed by integrating the longitudinal autocorrelation of the velocity fluctuations and the dissipation rate $\varepsilon$ is estimated via $\varepsilon = 0.7 \left| \mathbf{u}' \right|^3 / L_{\rm int}$ \citep{vassilicos_dissipation_2015}. Finally, the integral Reynolds number is $Re = \left| \mathbf{u}' \right| L_{\rm int} / \nu_a$, with $\nu_a$ the kinematic viscosity of air. The turbulent properties are summed up in Tab. \ref{tab:turb}, the values reported are the averages over the 5~cm above the surface of the bath

\begin{table*}
    \centering
    \begin{tabular}{cccccc}
        RPM & $\left| \left< \mathbf{U} \right> \right|$ (cm/s) & $\left| \mathbf{u}' \right|$ (cm/s) & $L_{\rm int}$ (cm) & $\varepsilon$ (m$^2$s$^{-3}$) & $Re$ \\ 
        0 & 0.0 & 0.0 & 2.3 & 0.0 & 0 \\
        50 & 2.0 & 2.0 & 1.2 & $4.0 \times 10^{-4}$ & 16 \\
        275 & 9.0 & 13.5 & 1.6 & 0.14 & 142 \\
        460 & 15.3 & 24.1 & 2.0 & 0.50 & 315 \\
        610 & 22.4 & 33.7 & 2.1 & 1.59 & 466 \\
    \end{tabular}
    \caption{Average values of the turbulence parameters.}
    \label{tab:turb}
\end{table*}

\section{Surface ageing and the need for overflow} \label{sec:overflow}

\begin{figure}
    \centering
    \includegraphics[width=0.8\linewidth]{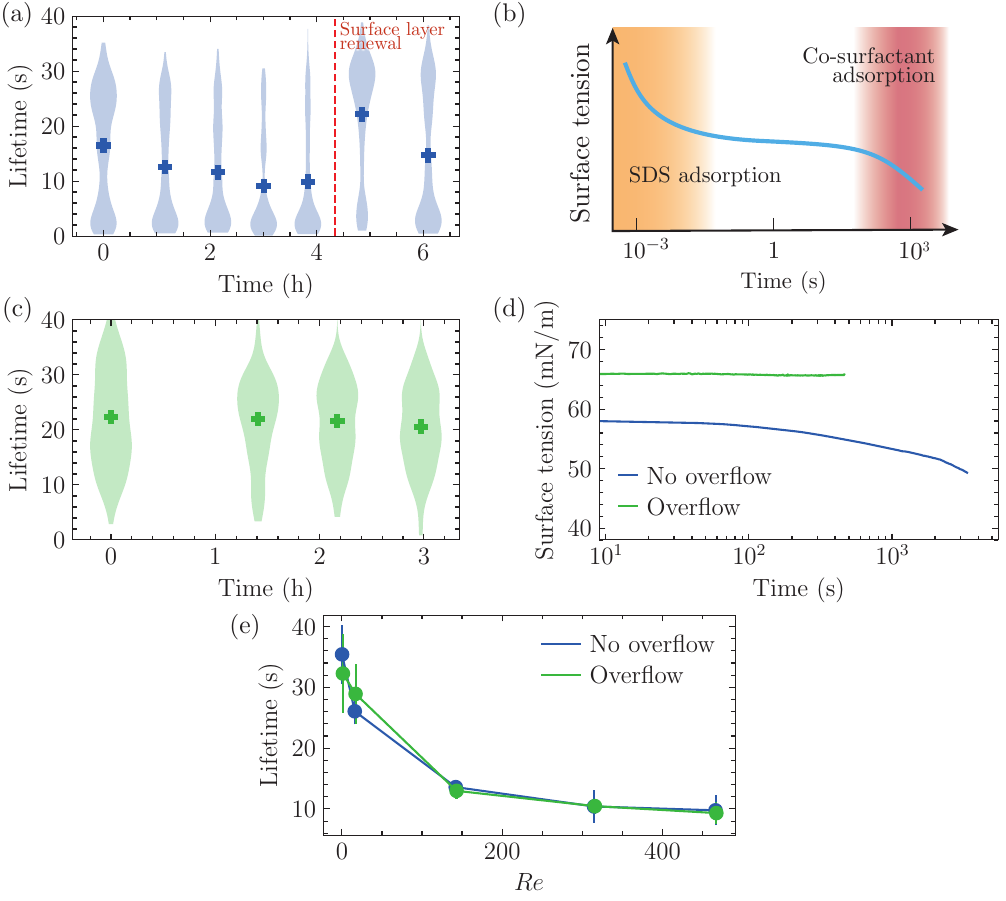}
    \caption{Effects of surface ageing on bubble lifetime. (a) and (c) show the distribution of lifetimes for several consecutive runs of 150 bubbles. The shaded area illustrates the whole distribution while the plus signs show the mean value. In (a) there is no overflow but after 4 hours (vertical red line) the surface layer is removed with a syringe ($c_{\rm SDS} = 200$ \textmu mol/L, $\mathcal{R}_H = 50\%$, no wind). In (c) the overflow is continuously turned on ($c_{\rm SDS} = 200$ \textmu mol/L, $\mathcal{R}_H = 50\%$, no wind, sea salt). (b) Schematic of the adsorption curve for SDS displaying two typical timescales: at early times SDS is adsorbing to the interface (orange area) and at late times the surface ages and the surface tension slowly decreases (red). (d) Measurements of the surface tension at long times with (green) and without (blue) continuous overflow of the solution. The surface layer is ageing and surface tension decreases in the case without overflow while it remains constant when the overflow is activated. (e) Lifetime as a function of the Reynolds number in the high concentration ($c_{SDS} \gtrsim$ CMC) limit where the effects of surface ageing are the least visible. The blue curve is the data with no overflow, while the green curve is the data with overflow. Vertical bars represent one standard deviation.}
    \label{fig:overflow}
\end{figure}

Surface properties need particular attention to be able to study bubble lifetime. Indeed, our goal in this study is to investigate the effect of various parameters on the mean lifetime of surface bubbles. In order to have robust conclusions however, we need to make sure that we can repeatably measure the mean lifetime for a given set of parameters. In particular, repeating the same experiment should yield the same result within some margin of error. This is not trivially achieved with surface bubbles as was previously discussed by \citet{poulain_ageing_2018}. It is not obvious in our case as well and illustrated in Fig. \ref{fig:overflow} (a) (using a solution containing 200 \textmu mol/L of SDS and no salt): while measuring the lifetime of 150 bubbles gives a good measurement of the mean (a shaded area illustrates the distribution for a group of 150 bubbles and a plus sign indicates the mean), repeating the same 150 bubble run reveals a slow drift of the mean lifetime towards short times. This drift induces lifetime differences that can be of tens of seconds, which is of the same order of magnitude as the differences we expect to see from changing the wind speed or the solution. Humidity and fan speed are continuously recorded and do not vary in time, temperature is also regularly checked and does not vary significantly. 

Our interpretation of this drift is the ageing of the surface layer and therefore a modification of its physical properties. This is made evident by the fact that, after four hours (or five runs) we removed the surface layer of the solution with a syringe (vertical red line). This removal results in a very clear change of the distribution, which becomes similar to the one of the first run. The drift towards short lifetime then resumes. The ageing of the surface can be caused by several effects: first, surface active molecules can be present in the air surrounding the setup in the experiment and slowly fall onto the surface (even though in our case the setup is in a closed box and the air filtered before entering). Similarly, surface active molecules can be found in small amounts because of imperfect cleaning of the experimental setup. Because of the steps taken to prevent these sources of impurities, we believe that these effects are small. Finally, it has been shown previously that even high purity surfactant products often contain remnants of other molecules used in the fabrication process that can have surface active properties such as dodecanol in the case of SDS \citep{fainerman_surface_2010,miguet_stability_2020}. These co-surfactants very slowly adsorb to the interface (over several hours) and strongly reduce the surface tension of the solution. One possibility is then to purify the surfactant used (as performed by \citet{miguet_stability_2020}) to remove the traces of other surfactants, however SDS also slowly undergoes hydrolysis to naturally form dodecanol, and it is likely that completely getting rid of co-surfactants from other sources for the entire duration of a series of experiments would be difficult.

Instead, we take advantage of large difference in timescale between the adsorption of SDS and all other co-surfactants. A sketch of the typical surface tension over time of an SDS solution (made without purifying the surfactant) can be found in Fig. \ref{fig:overflow} (b) based on measurements of the same curve in \citet{fainerman_surface_2010}. Depending on the concentration, the SDS adsorbs to the surface in a typical timescale of milliseconds to at most tenths of seconds (orange area), after that, the surface tension is relatively constant until the co-surfactants have significantly adsorbed to the interface, typically in thousands of seconds (red area). By continuously renewing the surface we can therefore obtain an interface that remains in this intermediate range where surface tension can be considered constant. Any new surface element quickly equilibrates with the rest of the surface and is then slowly advected towards a designated location on the edge of the tank, therefore not ageing for more than a minute. Experimentally, this is done by continuously overflowing the container where bubbles are generated into a larger container below using a small peristaltic pump (see Fig. \ref{fig:setup}). This overflowing technique has been used in previous experiments on bubble lifetime \citep{detwiler_aging_1978,struthwolf_residence_1984}, but its effects have never been analyzed quantitatively.

We have measured the static surface tension of SDS solutions using the Wilhelmy plate method with and without overflowing for timescales larger than 10 seconds and see that the drift is indeed prevented by overflowing the solution (shown in Fig \ref{fig:overflow} (d), the minimal time resolution our setup allows is 1~s). The result of implementing this approach in our setup can be seen in Fig \ref{fig:overflow} (c) where we repeat runs of 150 bubbles as in (a) but with the overflow activated. In contrast to the previous case, the lifetime distribution remains statistically similar for several hours. It should be noted that we have checked that the pump flow rate (from 0.5 to 1.7 mL/s) and therefore the renewal rate of surface elements does not affect the lifetime measurably and that this result is not specific to the SDS concentration (200 \textmu mol/L) used here. Regular renewal of the surface layer, in opposition to a still interface over a motionless subphase is also a reasonable working assumption when considering bubbles in the ocean or in rivers where there is turbulence close to the surface. In the following we always use a flow rate of 0.75 mL/s.

At high concentrations (around the CMC, and static surface tensions of about 30~mN/m), the effect of surface ageing is less visible because the effects of SDS are already very important. This allows us to compare lifetimes obtained in the cases with or without overflow (Fig \ref{fig:overflow} (e)). In this limit, both approaches yield similar results, validating the use of overflow to study the lifetime of surface bubbles. The same comparison is difficult to make at lower concentrations given large fluctuations in the lifetime without overflow. In practice, we never use the same solution for more than two consecutive days, and we estimate the minimal uncertainty on the mean lifetime measurement to be about one second.

\section{Results}\label{sec:results}

\begin{figure}
    \centering
    \includegraphics[width=0.9\linewidth]{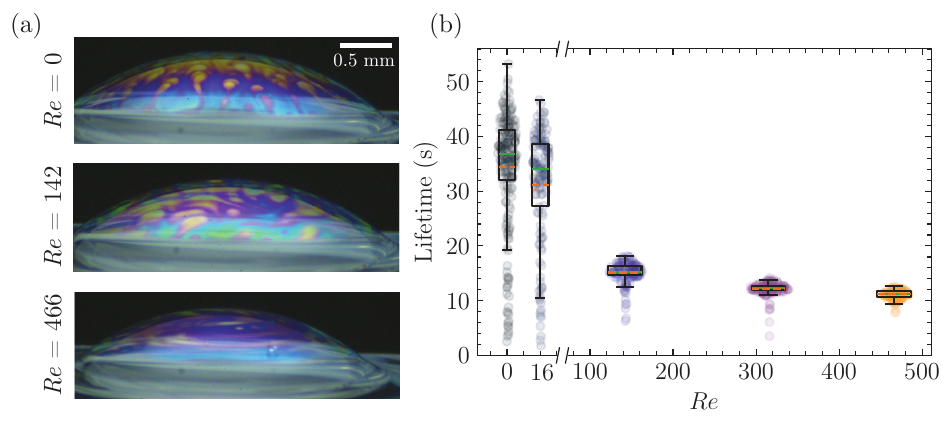}
    \caption{(a) Illustrations of thickness variations in the bubble cap using color interferometry. From top to bottom the Reynolds number is 0, 142, and 466. A supplementary movie showing the motion of the patches is available. (b) Lifetime as a function of the integral Reynolds number. Each circle represents a single bubble lifetime, the boxplot adds additional statistical information: (green line) median, (orange dashed line) mean, (box width) interquartile range, (whiskers) furthest data point within 1.5 times the interquartile range from the box. The data corresponding to a fan RPM of 50 or $Re = 16$ has been slightly shifted for clarity. The data presented is obtained with 200~\textmu mol/L of SDS and at a relative humidity of 50~\%.}
    \label{fig:quali}
\end{figure}

\subsection{Qualitative observations}

The presence of wind strongly disrupts the classical picture of the motion of the patches in the bubble cap (Illustrated with three panels of Fig. \ref{fig:quali} (a), see also the supplementary movie). In these frames white light creates interference patterns within the bubble cap and provides a view of thickness variations. In the top panel (no wind) we see patches generated at the foot of the bubble through marginal regeneration and rising in the cap. These patches allow the exchange of fluid between the bubble cap and the bath and are responsible for the drainage of the cap. They are generated through an instability of the pinch at the foot of the bubble \citep{lhuissier_bursting_2012,gros_marginal_2021,tregouet_instability_2021}. These patches are thinner than the rest of the film and therefore buoyant, and they detach from the pinch at the foot when they are large enough. Each patch generated therefore corresponds to a loss of volume in the bubble cap, and it has been shown that the volume change due to generation of these patches exactly matches the viscous limited flow out of the cap \citep{miguet_marginal_2021}. The motion of these patches in the cap can also be taken into account to model flows induced by surface tension gradients either due to evaporation or heating for instance \citep{poulain_ageing_2018}.

This picture is largely modified by the presence of wind (middle and bottom panel, and supplementary movie). Instead of rising straight up because of buoyancy, the patches are entrained by the wind in the air close to the cap. They are therefore constantly being deformed and pushed around by the strongly intermittent wind close to the bubble. In the middle panel we can still discern the patches, but some of them have a mostly horizontal velocity instead of vertical and their shape is quite different from the classical picture. In the bottom panel (largest wind forcing) the cap is almost completely mixed with the patches now taking the form of horizontal bands in the cap. In the case with the lowest wind forcing ($Re = 16$, not shown), the motion of the patches varies between being very similar to the no wind case and being entrained by the wind as an eddy reaches the proximity of the bath surface. We have verified that the presence or absence of the overflow has no significant effect on the bubble cap images or the conclusions above.

These observations could have important consequences for the drainage of the bubble as the size selection of the patches occurring at the foot of the bubble is most likely dominated by wind forcing instead of buoyancy for stronger wind speeds. As the patches are setting the drainage rate of the bubble, this could strongly alter the drainage profiles. However, as we show later in Fig. \ref{fig:drain}, the effect of the wind is important on the evaporative rate but not on marginal regeneration or capillary pressure induced drainage. 

Wind strongly reduces the mean lifetime of surface bubbles in solutions of pure surfactants. For a given solution and relative humidity, the lifetime can vary by a factor up to 5 only due to the intensity of the wind forcing. Fig. \ref{fig:quali} (b) illustrates this effect, (with 200 \textmu mol/L of SDS and a relative humidity of 50\%) where we can see several effects of the presence of wind: each point is an individual bubble lifetime measurement and the box plot shows the most important metrics of the distribution. The mean lifetime (orange horizontal lines) monotonously decreases as the Reynolds number increases, and so does the spread around the mean.  The reduction in the mean lifetime may \emph{a priori} be due to many effects: early bursting due to a perturbation from the wind, increased drainage through direct evaporation from the cap or increased flow through the bubble foot because of the strong cap mixing for instance. We note here that \citet{burger_persistence_1983} found an opposite trend of increasing lifetime when increasing wind over the surface with a limited dataset. The reason for this opposite trend is unknown. In the following we derive a drainage model taking into account only direct evaporation and show that it is sufficient to explain most of the observed variation of lifetime.

\subsection{Drainage}

We model the effect of the wind on bubble lifetime by taking into account its effect on the evaporative rate. The effect of the wind-induced motion in the cap is not considered in the following. There are two mechanisms responsible for the drainage of the cap of the bubbles: viscosity limited flow from the cap towards the bath due to the capillary pressure through the foot of the bubble and direct evaporation from the cap. Following the derivation from previous studies \citep{lhuissier_bursting_2012,poulain_ageing_2018,miguet_stability_2020}, the velocity through the foot of the bubble reads $u = \alpha_d (\gamma / \mu) (h / R)^{3/2}$, with $\gamma$ and $\mu$ the surface tension and viscosity of the liquid, respectively. $\alpha_d$ is a non-dimensional constant of order unity. The evaporative flux $j_e$ for an arbitrary convection in the air can be written as $j_e = \mathrm{Sh} j_{\rm diff}$, with $j_{\rm diff}$ the flux due to diffusion alone, and $\mathrm{Sh}$ the Sherwood number quantifying the importance of convective evaporation compared to diffusion alone. Replacing $j_{\rm diff}$ with its expression gives:
\begin{equation}
    j_e = \mathrm{Sh} \frac{D c_{\rm sat} (1 - \mathcal{R}_H)}{R \rho},
\end{equation}
where $D$ is the diffusivity of water vapor in air, $c_{\rm sat}$ the mass saturation concentration, $\rho$ is the density of the liquid.
Mass conservation in the cap of the bubble reads: 
\begin{equation}
    S\frac{\mathrm{d}h}{\mathrm{d}t} + Phu + S j_e = 0,
\end{equation}
where $h$ is the thickness of the cap, and $S$ and $P$ are the surface and perimeter of the bubble, respectively. Substituting the values of $u$ and $j_e$ and using $P / S \sim \ell_c / R^2$, with $\ell_c = \sqrt{\gamma / \rho g}$ the capillary length \citep{lhuissier_bursting_2012}, we obtain the equation describing the drainage of the cap:
\begin{equation}\label{eq:drain_dim}
    \frac{\mathrm{d}h}{\mathrm{d}t} + \alpha_d \frac{\gamma \ell_c}{\mu} \frac{h^{5/2}}{R^{7/2}} + \mathrm{Sh} \frac{D c_{\rm sat} \left(1 - \mathcal{R}_H\right)}{R \rho} = 0.
\end{equation}

The value of the Sherwood number depends on the Reynolds number in the air and the relative humidity. It can be divided into natural and forced convection depending on the value of the Reynolds number. At low Reynolds numbers (i.e. the two lower wind cases of our study), the evaporation is dominated by natural convection. In that case, the evaporation of the surface of the bath creates a plume of lighter high humidity air that rises in the center of the bath. The relevant dimensionless parameter in this problem is the Grashof number comparing evaporation-induced buoyancy and viscous dissipation: 
\begin{equation}
    Gr = \left| \frac{\rho_a(\mathcal{R}_H=100\%) - \rho_a(\mathcal{R}_H)}{\rho_a(\mathcal{R}_H)} \right| \frac{g \mathcal{L}^3}{\nu_a^2},
\end{equation}
with $\rho_a(\mathcal{R}_H)$ the density of the air, $\mathcal{L}$ a typical length scale of the bath surface and $\nu_a$ the kinematic viscosity of air. The Grashof number can be simplified into $Gr \approx (1 - \mathcal{R}_H) Gr_0$ with $Gr_0$ the maximal value of the Grashof number, occurring with a perfectly dry air in the chamber \citep{auregan_drainage_2025}. In our setup the Grashof number is typically $10^4$. \citep{dollet_natural_2017} have derived an expression for the flux over the surface of the bath: $j_e = D c_{\rm sat}(1 - \mathcal{R}_H) Gr^{1/5} / \mathcal{L}$. Assuming that the flux from the bubble cap is the same one as from the bath's surface, we obtain the Sherwood number associated with natural convection:
\begin{equation}
    \mathrm{Sh}_N \propto \frac{R}{\mathcal{L}} Gr^{1/5}.
\end{equation}
Note that in deriving the above equation, we have neglected the effects from the edge of the bath or the central plume similarly to \citet{miguet_stability_2020}.

At higher Reynolds numbers, the airflow above the bath is turbulent and dominates the evaporation. The rate at which bubbles evaporate is limited by the rate of the renewal of dry air elements close to the surface, which depends on the Reynolds number. The exact form of the relationship between the Sherwood and the Reynolds number depends on the properties of the flow, the geometry of the problem and can also depend on the interfacial properties \citep{theofanous_turbulent_1976,turney_air_2013,wissink_effect_2017}. This relation can be found by using Higbie penetration theory \citep{higbie_rate_1935,farsoiya_direct_2023}: the mass transfer velocity $k_e$ is given by $k_e \propto \sqrt{D / \tau}$, with $\tau$ the timescale at which air surface elements are renewed. The mass transfer velocity is then linked to the evaporation flux in our case through $j_e = k_e c_{\rm sat} (1 - \mathcal{R}_H) / \rho$, and therefore the Sherwood number associated with forced convection is $\mathrm{Sh}_F = k_e R / D$.
If the relevant timescale for surface renewal is the integral timescale $\tau_{\rm int} = L_{\rm int} / \left|\mathbf{u}'\right|$, then the Sherwood number reads:
\begin{equation}
    \mathrm{Sh}_F \propto \frac{R}{L_{\rm int}} Sc^{1/2} Re^{1/2},
\end{equation}
with $Sc = \nu_a / D$ the Schmidt number. If however, the relevant timescale is given by the smallest scales of the turbulent motion, then $\tau \sim (\nu_a / \varepsilon)^{1/4}$, and we get:
\begin{equation}
    \mathrm{Sh}_F \propto \frac{R}{L_{\rm int}} Sc^{1/2} Re^{3/4}.
\end{equation}

In the following, we use a Sherwood number combining the effects of natural and forced evaporation. The forced convection scaling based on the integral timescale is expected to be valid at low Reynolds numbers, while the one based on the small turbulent scales is expected to be valid at higher Reynolds numbers \citep{theofanous_turbulent_1976}. We will therefore assume that the relevant timescale is the integral timescale. 

Finally, to continuously cross from a purely convective regime at $Re = 0$ to a regime dominated by forced convection at high $Re$, we make the hypothesis that the total Sherwood number is the of sum the two contributions, with an added numerical prefactor $\alpha_e$ of order unity as the expressions above are only scaling laws. As a consequence, the Sherwood number can be written as the sum of a term dependent on $1 - \mathcal{R}_H$, and term dependent on the Reynolds number: 
\begin{equation}\label{eq:sh_definition}
    \mathrm{Sh} = \alpha_e \left(\mathrm{Sh}_N + \mathrm{Sh}_F\right) = \alpha_e \left(\frac{R}{\mathcal{L}}Gr^{1/5} + \frac{R}{L_{\rm int}}Sc^{1/2} Re^{1/2}\right).
\end{equation}
Typical values are 0.3 to 0.5 for $\mathrm{Sh}_N$ and 0.5 to 2 for $\mathrm{Sh}_F$, we also use the average value $L_{\rm int} = 2$~cm in the following.

We obtain the characteristic length and time scales of the problem by noticing that at early times drainage through the foot of the bubble dominates, allowing us to recover the classical drainage law for surface bubbles: $h \propto  (\gamma \ell_c t / (\mu R^{7 / 2}))^{-2/3}$ \citep{lhuissier_bursting_2012}. Evaporation becomes the main draining mechanism when the two draining terms are of the same order of magnitude: $\alpha_d \gamma \ell_c h^{5/2} / (\mu R^{7/2}) = \mathrm{Sh} \left(1 - \mathcal{R}_H\right) D c_{\rm sat} / (R \rho)$. This equation allows us to introduce the typical thickness where evaporation becomes the main draining mechanism $h^\star$, and using the early time drainage law, the associated typical timescale $t_c$:
\begin{equation}\label{eq:scales}
    h^\star = R \left( \frac{\mu D c_{\rm sat}}{\rho \gamma \ell_c} \right)^{2 / 5}\, {\rm and }\, t_c = \left(\frac{R^2 \rho}{D c_{\rm sat}}\right)^{3/5} \left( \frac{\mu R^2}{\gamma \ell_c} \right)^{2/5}
\end{equation}

We finally obtain the dimensionless drainage equation by normalizing Eq. \eqref{eq:drain_dim} using $\tilde{h} = h / h^\star$ and $\tilde{t} = t / t_c$:
\begin{equation}\label{eq:drain}
    \frac{\mathrm{d} \tilde{h}}{\mathrm{d}\tilde{t}} + \alpha_d \tilde{h}^{5/2} + \alpha_e \tilde{j_e} = 0,
\end{equation}
with $\tilde{j_e} = \left(\mathrm{Sh}_N + \mathrm{Sh}_F\right) \left(1 - \mathcal{R}_H\right)$ the dimensionless evaporative flux. This model is similar to the one employed by \citet{poulain_biosurfactants_2018} with the difference that we explicitly estimate the dependence of evaporative rate with wind speed and humidity.

\begin{figure}
    \centering
    \includegraphics[width=0.6\linewidth]{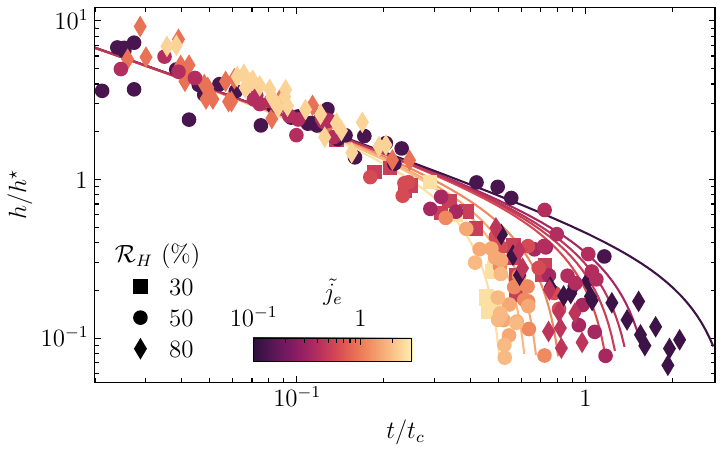}
    \caption{Cap thickness as a function of time in dimensionless units. Markers indicate relative humidity and color shows the dimensionless evaporative rate $\tilde{j_e} = \left(\mathrm{Sh}_N + \mathrm{Sh}_F\right) \left(1 - \mathcal{R}_H\right)$. Full lines are solutions of Eq. \eqref{eq:drain}, with $\alpha_d=1.8$ and $\alpha_e=1.2$. The data shown was obtained with a solution of 200 \textmu mol/L of SDS.}
    \label{fig:drain}
\end{figure}

We validate this model only taking into account wind speed through evaporation using experimental data. We analyze the effect of humidity and wind on bubble drainage by measuring the thickness of the cap over time. To do so, we record the spontaneous bursting of surface bubbles using a high speed camera. This allows us to track the retraction of the rim as a hole opens in the cap. The thickness is then accessible using the classical Taylor-Culick retraction velocity: $V_\textsc{TC} = \sqrt{2 \gamma / (\rho h)}$ \citep{lhuissier_bursting_2012,poulain_ageing_2018,shaw_film_2024}. We repeat this process for several humidity values and for about 20 bubbles each time. 

The result is shown in Fig \ref{fig:drain} using an SDS solution (200 \textmu mol/L): the thickness is normalized by the units defined in Eq. \eqref{eq:scales}. Different markers correspond to different relative humidities and color indicates the dimensionless evaporative rate $\tilde{j_e} = \mathrm{Sh} \left(1 - \mathcal{R}_H\right)$. For $t / t_c \lesssim 0.4$, we find that all the data for different evaporative rates (humidity or wind speed) indeed collapse on the same trend, close to $h \propto t^{-2/3}$.  Finding that this classical result is unchanged shows that the strong mixing of the cap at high wind speed actually has little effect on the drainage of the bubbles. Further, the motion of the marginal regeneration patches (their rise velocity or the size at which they detach from the foot of the bubble) does not seem to have a noticeable effect on the drainage in this case.

After a time which is typically of order $0.4\times t_c$, we see the cases at different evaporative rates separate and the drainage becomes faster. The thickness now decreases linearly with time, with the cases with strong evaporation (yellow) decreasing faster and earlier than the cases with little to no evaporation (black). As evaporation becomes dominant, the location where the bursting event is initiated goes from being homogeneously distributed in the cap to being more likely close to the top (see App \ref{app:burst_angle}).

We can quantitatively predict the drainage behavior by integrating Eq. \eqref{eq:drain} over time combined with the Sherwood number given by Eq.~\eqref{eq:sh_definition} by fitting the numerical prefactors $\alpha_d$ and $\alpha_e$ on the data. Note that in order to do so, we need to estimate the thickness at generation $h_0$. We estimated this value by using an average over the earliest bursting events we recorded (with a lifetime shorter than 1 second), and found this initial thickness to be independent of wind speed and humidity and equal to $h_0 = 12$ \textmu m. All other physical parameters of the model are either measured or computed from the properties of the solution. We see a good agreement with our measurements, further validating the fact that the main draining mechanisms are i) viscous limited flow through the foot of the bubble and ii) direct evaporation from the cap. The numerical value of the prefactors we obtained are $\alpha_d = 1.8$ and $\alpha_e = 1.2$ for the whole dataset. The numerical integration slightly overshoots the cases at high humidity and low wind, possibly because of the hypothesis made in estimating the natural convection Sherwood number. 

%The drainage picture is fully consistent with the lifetime approach detailed above: for instance setting a bursting criterion at a very small thickness ($h_c < 10^{-3}h_0$) allows us to numerically recover the result $t_f \sim t_c$. 

\subsection{Lifetime of solutions of pure surfactants}

\begin{figure}[htbp]
    \centering
    \includegraphics[width=0.85\linewidth]{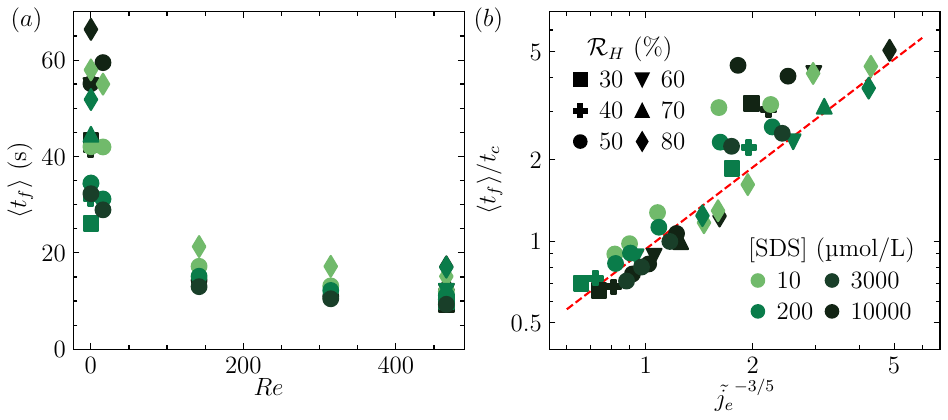}
    \caption{{a Mean lifetime as a function of the Reynolds number $Re$, for various relative humidity and surfactant concentration. (b) Dimensionless lifetime as a function of $\tilde{j_e}^{-3/5}$. Markers show relative humidity and colors show SDS concentration. (red dashed line) Eq. \eqref{eq:tfmax}: $t_{f, {\rm max}}=1.32 \alpha_e^{-3/5} \alpha_d^{-2/5} \times \tilde{j_e}^{-3/5}$, with $\alpha_d=1.8$ and $\alpha_e=1.2$.}}
    \label{fig:lifetime_sds}
\end{figure}

In this section, we analyze the lifetime $t_f$ of surface bubbles (from the instant the bubble reaches the surface until it spontaneously bursts) in a solution only containing SDS (no salts). Plotting our data for several wind speeds, relative humidities and SDS concentrations as a function of the Reynolds number (Fig. \ref{fig:lifetime_sds} (a)) shows a behavior similar to the one observed in Fig. \ref{fig:quali} (b), for all SDS concentrations (color) and humidities (markers). The mean lifetime with little to no wind ranges from 30 to 60 seconds and decreases down to 10 seconds for all cases as the Reynolds number increases.

We can compute the maximal time $t_{f, {\rm max}}$ that one can expect a bubble to remain on the surface before bursting using Eq. \eqref{eq:drain}. Following the derivation in \citet{poulain_biosurfactants_2018}, in the limit when the bursting thickness $h_c$ is very small compared to the thickness at generation ($h_c \ll h_0$), this time is given by:
\begin{equation}
    t_{f, {\rm max}} = t_c \int_{\tilde{h_c}}^{\tilde{h_0}}\frac{\mathrm{d}\tilde{h}}{\alpha_e \tilde{j_e} + \alpha_d\tilde{h}^{5/2}} \approx t_c \left( \alpha_e \tilde{j_e}\right)^{-3/5} \alpha_d^{-2/5} \int_0^\infty \frac{\mathrm{d}\tilde{h}}{1 + \tilde{h}^{5/2}}, 
\end{equation}
and the last integral can be evaluated to obtain:
\begin{equation}\label{eq:tfmax}
    t_{f, {\rm max}} \approx 1.32 \:t_c \left( \alpha_e \tilde{j_e}\right)^{-3/5} \alpha_d^{-2/5}.
\end{equation}
We therefore plot our mean lifetime data (normalized by $t_c$) as a function of $j_e^{-3/5}$ in Fig. \ref{fig:lifetime_sds} (b) and the data indeed follows a linear trend. This graph reveals that most of the changes in mean lifetime $\left< t_f \right>$ are captured using a single parameter that contains the effects of wind and humidity on the evaporative rate: $\tilde{j_e} = \left(\mathrm{Sh}_N + \mathrm{Sh}_F\right) \left(1 - \mathcal{R}_H\right)$. This result confirms that our modelling of the Sherwood number in the large range of conditions (from quiescent to fully turbulent) is correct. In addition, it shows that there are no visible effects of the wind on bubble lifetime beyond altering the evaporative rate. In this case with no salts in the solution, the effects of the surfactant concentration are small and do not show a significant trend.

The maximal lifetime $t_{f, {\rm max}}$ computed from Eq. \eqref{eq:tfmax} (red dashed line) follows closely the data and in particular is in quantitative agreement with our experimental measurements of $\left< t_f \right>$. This last result is consistent with the fact that the distribution of lifetimes is narrow around its mean (see Sec. \ref{sec:distribution}) such that the maximal and mean lifetime are of the same order of magnitude.
We note that in some of the cases, the mean experimental lifetime is larger than the maximal expected lifetime. This discrepancy may be due to small experimental differences between the drainage and lifetime setups: the tank in which the drainage experiment is performed is smaller (3~cm in diameter instead of 7~cm) such that the bubble always remains in view of the high speed camera. The proximity of a larger bath decreases the evaporative rate and therefore increases the maximal possible lifetime. At very high concentrations (black markers) other possible effects include a reduced evaporation rate due to the extreme contamination of the interface \citep{takemura_rising_1999} or a slowed down drainage due to the rigidification of the interface.

\citet{poulain_biosurfactants_2018} observed for bubbles in the large surfactant concentration limit, that the longest time a surface bubble can live is given by $t_{f, {\rm max}}$. With our setup in which we took great care in removing impurities (see Sec \ref{sec:overflow}), we find that the average lifetime is of order $t_{f, {\rm max}}$ (instead of just the maximum). This result is valid for very large changes of SDS concentration, from values greater than the CMC down to just a few \textmu mol/L (or $10^{-3}$ times the CMC). All of our lifetime data for SDS is well collapsed when plotted as $\left< t_f \right> / t_c = f(j_e^{-3/5})$. This result is particularly robust at high wind speeds (or low $j_e^{-3/5}$), which is coherent with our previous observations that mixing the air around the bath and the bubble improves the repeatability of lifetime experiment by preventing local fluctuations of the humidity field. At high $j_e^{-3/5}$, we observe lifetimes that can be larger than the trend, and in general more important fluctuations. These may be due to the difficulty in modeling the Sherwood number associated with natural convection.

\subsection{Lifetime of solutions containing salts}

\begin{figure}
    \centering
    \includegraphics[width=0.9\linewidth]{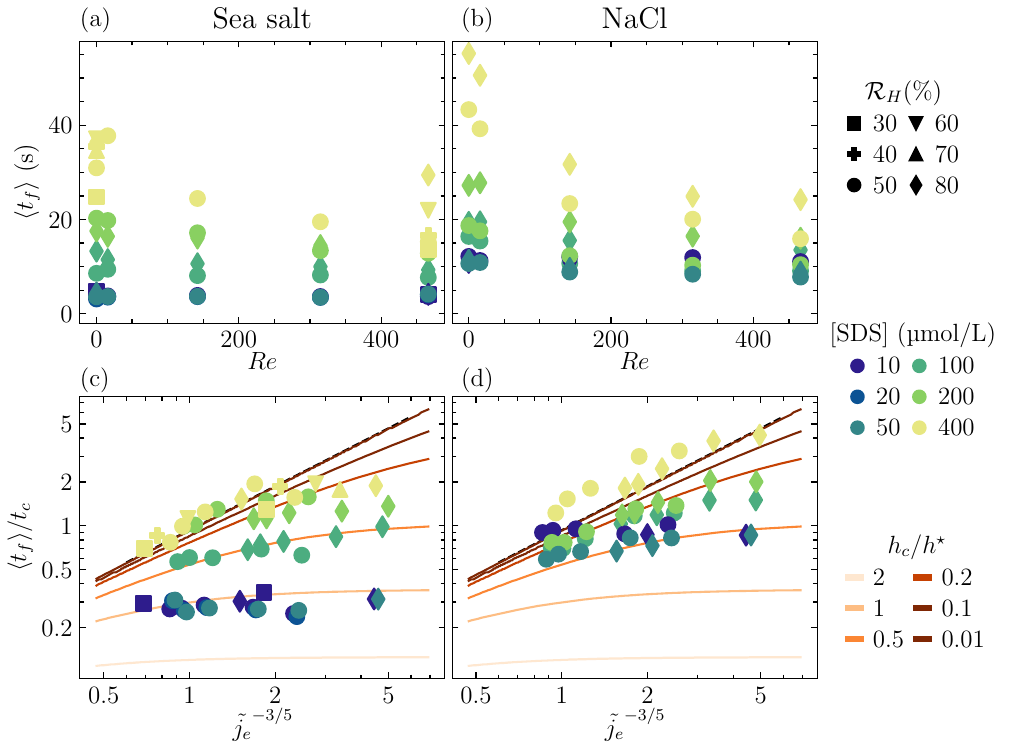}
    \caption{(a) and (b) Mean lifetime as function of $Re$, for various relative humidity, surfactant concentration, for the sea salt and NaCl cases, respectively. (c) and (d) Dimensionless lifetime as a function of $\tilde{j_e}^{-3/5}$. The left column shows the dataset with sea salt and the right one with NaCl. The dashed lines represent $t_{f,{\rm max}}$ and are identical to Fig. \ref{fig:lifetime_sds} (c) , given by Eq. \eqref{eq:tfmax}. The orange lines are obtained by integrating of Eq. \eqref{eq:drain} until the thickness reaches a critical thickness $h_c$. Markers show relative humidity and color shows SDS concentration.}
    \label{fig:lifetime_salt}
\end{figure}

The lifetime of bubbles in solutions of salt and surfactants is drastically different from that of bubbles in pure water and surfactants. The lifetime data is summed up in Fig. \ref{fig:lifetime_salt} for sea salt (a mixture of salts and inorganic compounds in their oceanic proportions) and NaCl only (35 g/L). The top row shows the dimensional mean lifetime as a function of $Re$ where we see that in cases with a higher SDS concentration (yellow) the behavior is similar to cases with SDS only, with a gradual decrease of the mean lifetime as $Re$ is increased. At low concentrations however ($c_\textsc{SDS} \lesssim 50$~\textmu mol/L, blue), the lifetime is essentially independent of wind speed or relative humidity. In this regime, the lifetime of solutions with sea salt is about 5~s, and 10~s with NaCl. It is interesting to note that in both cases, in this limit the lifetime is again independent on the surfactant concentration within the range tested. Even though the concentrations in SDS are fairly small (less than 5\% of the CMC), the combined presence of salts and surfactants results in a large variation of the surface tension (see Appendix \ref{app:lang}). 

The bottom row shows the same data as a function of $\tilde{j_e}^{-3 / 5}$ illustrating that in this case too, the dimensionless evaporative rate is an adequate parameter to combine the effects of changing wind speeds and relative humidities. We can now see in both cases a transition between a regime where bubble lifetime is much shorter than $t_{f, {\rm max}}$ (dashed black line) and independent of relative humidity or wind speed, and a regime again dominated by evaporation, with the lifetime given by Eq. \eqref{eq:tfmax}, as in Fig. \ref{fig:lifetime_sds} for pure water and SDS. The transition occurs for SDS concentration of around 100 and 200~\textmu mol/L, where the increase of lifetime with $\tilde{j_e}^{-3 / 5}$ is slower than predicted by Eq. \eqref{eq:tfmax} and the lifetime always remains inferior to the pure water and SDS case (dashed line). For the highest concentrations, the lifetime follows this trend again, showing that the drainage has reached an evaporation dominated regime. 

The case with NaCl and the largest amount of surfactant is noticeably above the rest of the data points, following the same linear trend but with a larger prefactor. It should be noted that in both cases this regime corresponds to an almost saturated isotherm (with a very low surface tension that does not vary with surface compression), similar to what would be obtained for a solution above the CMC. Finally, in the case of sea salt, the lifetime data departs from the trend at higher $\tilde{j_e}^{-3 / 5}$ (high relative humidity or low wind speed) even at the highest SDS concentration, and increases slowly with $\tilde{j_e}^{-3 / 5}$.

The transition observed for surface bubbles with salts can qualitatively be replicated numerically by integrating Eq. \eqref{eq:drain} until the thickness reaches a finite thickness $h_c$. We make here the hypothesis that the drainage curves obtained in Fig. \ref{fig:drain} remain unchanged upon the addition of salts and therefore again use the same prefactors $\alpha_d = 1.8$ and $\alpha_e=1.2$. This approximation is supported by results from \citet{shaw_film_2024} who showed that the effect of sea salt on drainage is much smaller than other effects (such as solution temperature for instance) and that the evolution of the thickness is still well captured by $h \sim t^{-2/3}$ at early times.

Results of this integration for various $h_c$ are shown in Fig. \ref{fig:lifetime_salt} in shades of orange. With decreasing $h_c$ we see a transition from a regime where the lifetime obtained is smaller than $t_{f, {\rm max}}$ and is independent of relative humidity or wind speed when $h_c \gtrsim h^\star$, to the evaporation dominated regime (obtained analytically) for $h_c \ll h^\star$ where the lifetime is given by $t_{f,{\rm max}}$. This transition qualitatively matches the one observed for sea salt or NaCl, showing that taking into account the effect of wind and relative humidity through the evaporative rate, and the effect of the solution through an average bursting thickness is enough to describe most of the mean lifetime trends seen in this study. 

The model predicts that the average bursting thickness in the wind independent regime is $h_c / h^\star \approx 1$ or $h_c \approx 0.6$~\textmu m for sea salt and $h_c / h^\star \approx 0.5$ or $h_c \approx 0.3$~\textmu m for NaCl. Both cases become evaporation dominated when $h_c$ is less than 100~nm. Values at rupture between $h_c \approx 0.3$ to $0.6$~\textmu m are coherent with previous measurements on the drainage and rupture of bubbles in the presence of salts \citep{shaw_film_2024}.

\section{Lifetime probability distribution}\label{sec:distribution}

\begin{figure}[htbp]
    \centering
    \includegraphics[width=\linewidth]{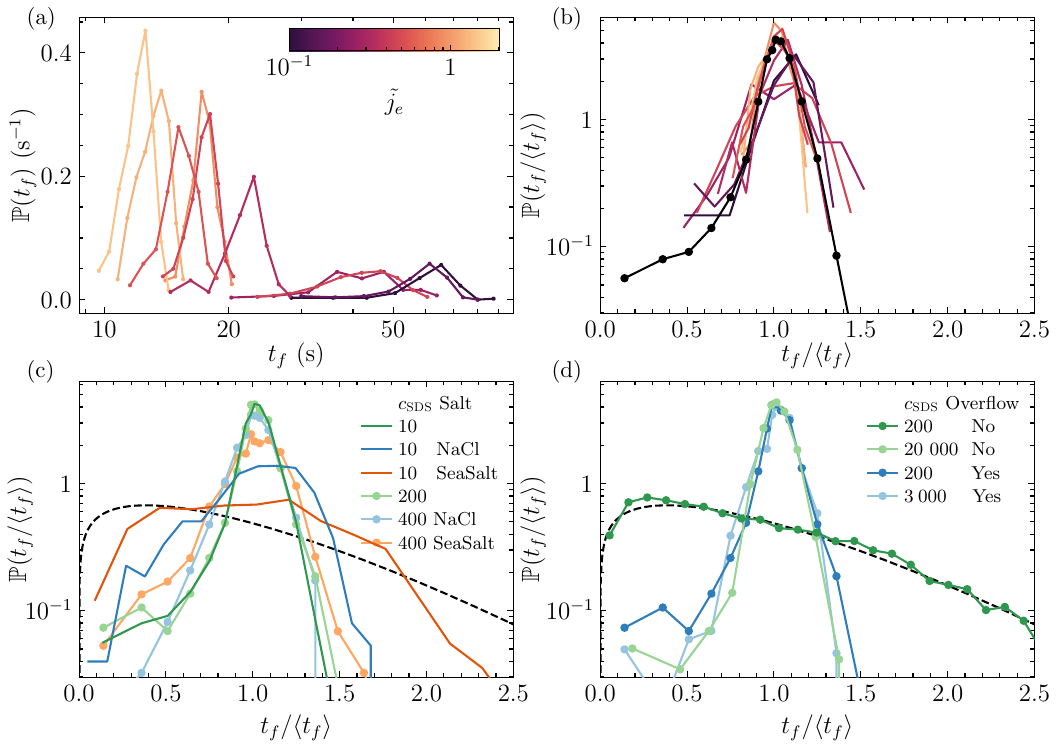}
    \caption{Distributions of lifetimes in various cases. (a) Dimensional distributions of lifetime for various evaporation rates (color) with a solution of 10~\textmu mol/L of SDS. (b) Same data, but each pdf is normalized by the mean lifetime, (black line with markers) Combined pdf of all cases. (c) Combined pdf for selected solutions, legend indicates $c_\textsc{SDS}$ in \textmu mol/L and the type of salt. (d) Comparisons of the combined pdf with and without overflow. In (b) and (c) the dashed black line is a Weibull distribution of shape $4/3$ and scale $\Gamma(7/4)^{-1}$.}
    \label{fig:distribution}
\end{figure}

In addition to studying the mean lifetime, we now turn our attention to the complementary approach of looking at the statistical distribution of lifetimes as it can inform us on the bursting mechanisms. To do so, we combine the data from multiple cases with a single solution but various humidity or wind speed to produce a single probability density function (PDF) per solution. Since the mean lifetimes for different evaporative fluxes (obtained by varying relative humidity and Reynolds) can be quite different (Fig. \ref{fig:distribution}, (a)), we combine the data by normalizing each run of around 150 bubbles with its average $\left< t_f \right>$. This approach is valid as, despite the pdfs in physical units $\mathbb{P}(t_f)$ being very different for different evaporative rates, the pdfs normalized by the means $\mathbb{P}(t_f / \left< t_f \right>)$ are remarkably similar (see Fig. \ref{fig:distribution}, (b), colored curves). As a result, the combined pdf (black curve with markers) is well-defined and robust as it contains several thousands single bubble bursting events. This approach of combining several cases with different lifetimes into a single normalized probability distribution was originally used in \citet{lhuissier_bursting_2012}.

Using this approach, we can clearly see different bursting mechanisms in action for different solutions. In the case of deionized water with only SDS and no surfactants, the distributions of lifetime are extremely narrow around the mean, regardless of the concentration (Fig. \ref{fig:distribution}, (c) green curves). This regime, as was described above, corresponds to a mean lifetime set by the evaporative rate, with a low likelihood of a bursting event before the time given by $t_{f,{\rm max}}$ (Eq. \ref{eq:tfmax}). We can also notice here that the width of the distribution is similar for low and high concentrations.

When salts are present (blue and orange curves), two regimes need to be distinguished: at low SDS concentrations, the addition of salt widens the distribution, but at high contamination, all cases are peaked with a similar shape as in the case without salt. Sea salt has a larger effect on the distribution than NaCl only. These observations are fully consistent with the mean lifetime approach of the rest of the study: the case without salt bursts at a time given by the evaporative rate, regardless of concentration. This results in a very narrow pdf around $t_{f, {\rm max}}$. For the cases with salts, at low SDS concentration the lifetime is independent of the evaporative rate and the bubbles can burst for a relatively wide range of lifetimes. At high concentration however the lifetime is dominated by evaporation, as seen from the mean lifetime, and the pdf is therefore also peaked. 

In order to describe the distributions of bubble lifetime \citet{lhuissier_bursting_2012} have derived the following expression: 
\begin{equation}\label{eq:weibull_lh12}
    \mathbb{P}\left(\frac{t_f}{\left< t_f \right>}\right) = \frac{4}{3} \Gamma\left(7/4\right)^{4/3}\left(\frac{t_f}{\left< t_f \right>}\right)^{1/3} \exp\left(-\left(\Gamma\left(7/4\right)\frac{t_f}{\left< t_f \right>}\right)^{4/3}\right),
\end{equation}
which is a Weibull distribution  of shape $4/3$ and scale $\Gamma(7/4)^{-1}$, with $\Gamma$ the gamma function. This parameter-less distribution has been successfully used to describe lifetime data in several studies \citep{poulain_ageing_2018,miguet_how_2020,lorenceau_lifetime_2020}. However, when plotted along our data (dashed black curve), we see that it does not describe any of the cases presented in this study: this distribution is extremely wide, and even in the cases with low SDS concentration and salt, the experimental distribution of lifetime is narrower around the mean than the proposed distribution Eq. \eqref{eq:weibull_lh12}.

To understand this discrepancy, we plot in panel (d) the distributions obtained when comparing similar cases with (blue) and without (green) overflow. At high SDS concentration (about the CMC), all cases are evaporation-dominated and therefore have a similarly narrow distribution. At lower concentrations however, the case with overflow remains very narrow while the case without overflow perfectly agrees with Eq. \eqref{eq:weibull_lh12}. The interpretation of the change in lifetime distribution is discussed below.

\section{Discussion}\label{sec:discussion}

The data presented in this article illustrates two very different regimes of surface bubble lifetime. For pure water and SDS only, the mean lifetime is essentially independent of the SDS concentration (from about one thousandth of the CMC to above the CMC) and closely follows the maximal possible lifetime given by Eq. \eqref{eq:tfmax}, with a peaked distribution. When either NaCl or sea salt is added, the mean lifetime is greatly reduced for a given SDS concentration and the distribution is wider. We then see a transition from being independent of wind speed at low SDS concentrations to being dominated by evaporation at high concentrations as in the case of pure water and SDS.

We believe the difference between these two regimes lies in the presence of impurities in the solution, which are required to trigger a bursting event at a large thickness. A large thickness means here bubble caps thicker than 1 \textmu m, in opposition to the very small thicknesses ($h \lesssim 0.1$~\textmu m) attainable for long-lasting bubbles. A bubble of pure water and surfactant does not burst before evaporation becomes the dominant drainage mechanism and rapidly (linearly in time) removes material from the cap. The regime where evaporation is dominant has previously been identified as the upper limit on bubble lifetime \citep{poulain_biosurfactants_2018}, the difference with our study is the use of the overflow to remove most co-surfactants from the interface and ensure a precise control of surface properties. Overflow seems an efficient way of also removing the small amount of impurities  (other surface active compounds than the main surfactant) that may be present when using only deionized water and surfactants in addition to ensuring that the physicochemical properties of solution remain constant (see Sec. \ref{sec:overflow}). 

Authors have previously hypothesized that surface tension inhomogeneities (for instance across the marginal regeneration patches) may be enough to trigger bursting of bubble cap \citep{lhuissier_bursting_2012}. From our data it seems that with SDS only, these inhomogeneities are not enough to make the bubble burst. This fact is supported by the observation that wind, which drastically alters the marginal regeneration patches and presumably any Marangoni flow that might be present within the bubble cap, does not seem to have an effect on lifetime or drainage beyond a change of the evaporative rate. 

Introducing salts changes the surface tension but also adds impurities to the solution. These impurities may then trigger large thickness bursting as theorized in \citet{poulain_ageing_2018} (Sec 6). The sea salt that we used contains many different salts but also other inorganic compounds and therefore more impurities than NaCl (which is 99\% pure). This fact may help explain why for a given SDS concentration, the sea salt case always has a shorter lifetime than the NaCl case (5 versus 10 s in the lower SDS concentration limit) and a broader distribution. 

Finally, another clue towards the need for impurities to trigger bursting at a large thickness is a few experiments we performed with tap water instead of deionized water (therefore also containing impurities). The resulting mean lifetime was close to 4~s with a small concentration of SDS (10 \textmu mol/L) in contrast to about 15~s with deionized water. We also performed experiments with Triton X-100, a non-ionic, slow adsorbing surfactant, which confirms the trends observed with SDS: for a given small concentration, the mean lifetime follows $\left< t_f \right> \sim t_{f, {\rm max}}$ with surfactant alone, but it is a constant at around 4~s when sea salt is added. These two test cases (shown in App. \ref{app:txtap}) point towards impurities being the reason for the difference between the cases with and without salts, instead of an electrostatic interaction specific to SDS and ions in solutions.

The observations on bubble lifetime distribution are also fully consistent with the interpretation that the presence of impurities in solution is responsible for the large thickness bursting regime. The cases with little impurities (with overflow and no salt) display a very peaked distribution around the mean, with a similar relative standard deviation across all cases. Introducing impurities widens the distribution by allowing the bubble to burst earlier than $t_{f, {\rm max}}$. Finally, the expression proposed by \citet{lhuissier_bursting_2012} can be though of as the limit at a large amount of impurities, where the cases without overflow aggregate. At very high surfactant concentration, the bubble cap is stable enough so that effect of impurities is less, and we recover peaked distributions and mean lifetimes given by the evaporative rate.
%Put the data somewhere ?

Knowledge of the surface tension isotherm of the solution considered is not enough to predict the expected lifetime of a surface bubble. Indeed, the amount of impurities cannot be directly inferred from the isotherm and therefore neither can the expected bursting thickness. A method of measuring the nature and concentration of impurities in a solution may therefore help in predicting the exact lifetime of a surface bubble.  \\

\noindent\textbf{Acknowledgement} We thank an anonymous reviewer for their insightful comments that have significantly improved the theoretical derivation in this manuscript. This work was supported by NSF grant 2242512, NSF CAREER 1844932 and the Cooperative Institute for Modeling the Earth's System at Princeton University to L.D.  \\
\noindent\textbf{Declaration of Interests.} The authors report no conflict of interest.

\appendix

\section{Langmuir trough measurements}\label{app:lang}

\begin{figure}
    \centering
    \includegraphics[width=0.9\linewidth]{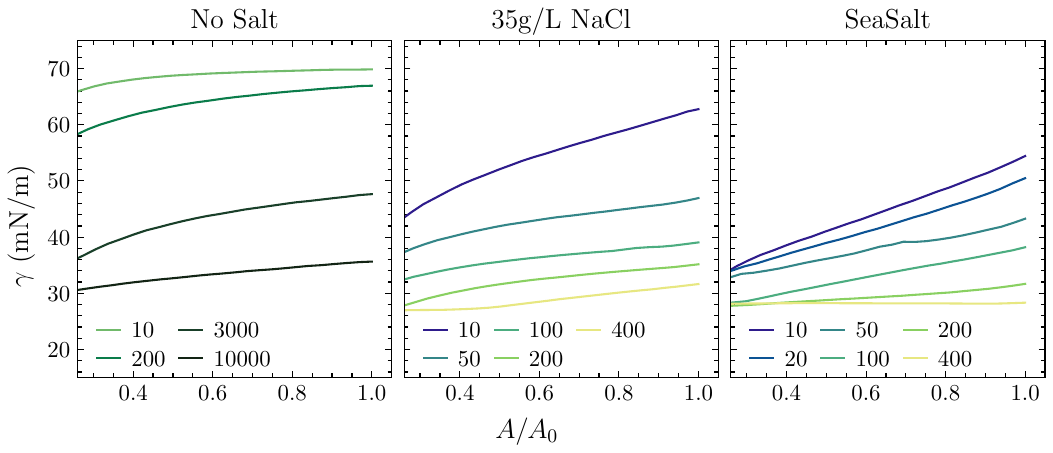}
    \caption{Surface tension isotherms of the various solutions used in this study. Numbers indicate the SDS concentration used in \textmu mol/L. }
    \label{fig:langmuir}
\end{figure}

Every solution used in this study was tested in a Langmuir trough (KSV NIMA, model KN 1003) using the Wilhelmy plate method to record its surface tension isotherm. We record the properties of the surface by setting aside a small amount of solution and placing it in the trough, and the surface tension is measured with a platinum Wilhelmy plate. After that, two Teflon barriers compress the surface at a rate of 270~mm/min allowing us to obtain the surface tension $\gamma$ as a function of the through area. We performed the measurement as quickly as possible after pouring the solution into the trough to avoid the adsorption of co-surfactants to the interface. The resulting data is plotted in Fig. \ref{fig:langmuir} as a function of the area, normalized by the initial area of the trough.

The range of concentrations tested with SDS only corresponds to the full range of interface properties from very clean (flat isotherm at 72~mN/m) to an almost fully surfactant packed interface (flat isotherm at 30~mN/m). The isotherms with salts also show a broad range of interface properties, but with a much narrower concentration range. Because of other surfactants present in the salts used and of the interaction between SDS and ions in the solution the maximal surface tension obtained in those cases is 63~mN/m with NaCl and 54~mN/m with sea salt. Both of these cases reach a fully packed case at 400~\textmu mol/L of SDS with a flat isotherm at about 30~mN/m.

\section{Other cases}\label{app:txtap}

\begin{figure}[htbp]
    \centering
    \includegraphics[width=0.6\linewidth]{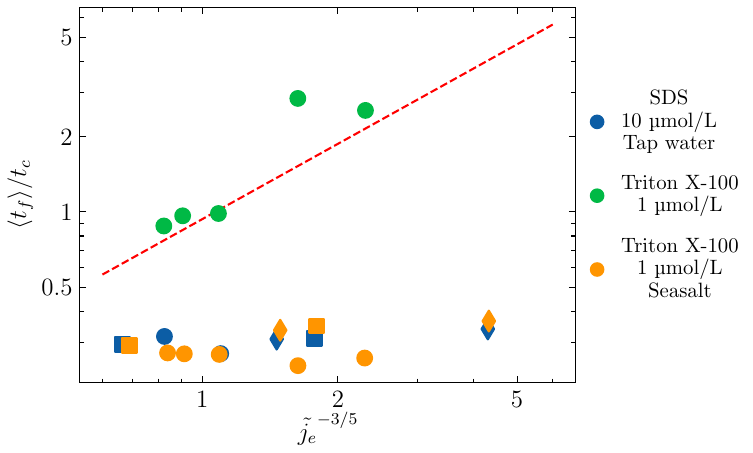}
    \caption{Dimensionless mean lifetime as a function of $\tilde{j_e}^{-3 / 5}$ for cases not shown in the main text: (blue) 10 \textmu mol/L of SDS mixed with tap water instead of deionized water, (green) 1 µmol/L of Triton X-100, and (orange) 1 µmol/L of Triton X-100 with sea salt added. The dashed red line is $1.32 \alpha_e^{-3/5} \alpha_d^{-2/5} \times \tilde{j_e}^{-3/5}$, with $\alpha_d=1.8$ and $\alpha_e=1.2$. Markers show relative humidity and are identical to Fig. \ref{fig:lifetime_salt}.}
    \label{fig:txtap}
\end{figure}

We ran a few test cases in addition to the ones presented in the main text. These tests include the use of Triton X-100, a non-ionic, slow adsorbing surfactant \citep{fainerman_adsorption_2009} to verify that the results presented here are not specific the SDS. We used 1 \textmu mol/L (note the CMC of Triton X-100 is 230 \textmu mol/L). These tests also include the use of tap water instead of deionized water (with SDS and no salt) to test if the introduction of impurities has the same effect as introducing salts. 

The results of these tests (Fig. \ref{fig:txtap}) are fully coherent with the discussion presented in the main text. First, the use of tap water results in a very low mean lifetime (around 3 to 4 seconds) that does not depend on the evaporation rate, in contrast to the case with deionized water (Fig. \ref{fig:lifetime_sds}) that scales linearly with $\tilde{j_e}^{-3/5}$. This effect is similar to adding NaCl or sea salt, and we can interpret it as the presence of impurities in the tap water that we use that can trigger an early bursting event. 

Second, the results with Triton X-100 are similar to the ones with SDS: without salt, the mean bubble lifetime is very close to the maximal lifetime given by $t_{\rm f, max}$ (Eq. \eqref{eq:tfmax}) while upon the addition of sea salt, the lifetime drops to around 3 to 4 seconds and is independent of the evaporation rate. The concentration of Triton X-100 used is smaller than in the case with SDS, but we expect that (i) similarly to the case with SDS in the absence of salt there is little dependence with surfactant concentration once the regime dominated by evaporation is reached and (ii) in terms of fraction of the CMC, the concentration of Triton X-100 (0.13\%) is close to the smallest SDS concentration used (0.43\%).

\section{Burst location}\label{app:burst_angle}

\begin{figure}[htbp]
    \centering
    \includegraphics[width=0.65\linewidth]{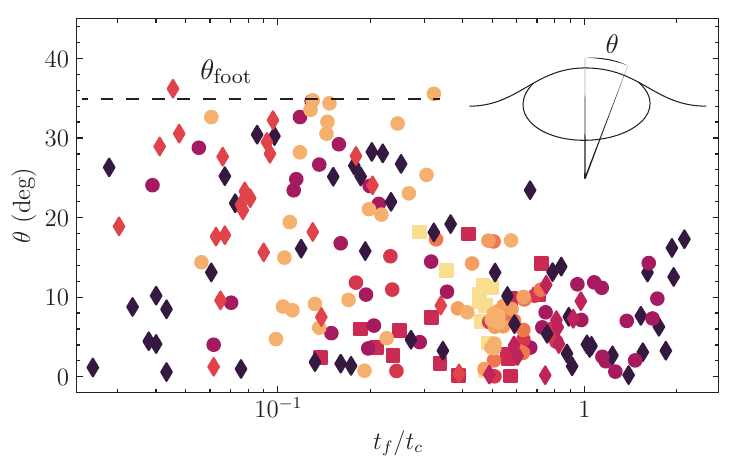}
    \caption{Angle of the location on the cap where the hole is initiated. $\theta = 0$ corresponds to the top of the cap, while $\theta \approx 35$~deg corresponds to the foot of the bubble. Colors represent the evaporation rate and markers show the relative humidity, identically to Fig. \ref{fig:drain}.}
    \label{fig:burst_angle}
\end{figure}

In addition to allowing us to measure the thickness through the film retraction, our high speed images of bubble rupture also allow us to locate the origin of film rupture. For each event we therefore recorded the angle $\theta$ on the spherical cap where the hole is initiated and plot it as a function of the lifetime in Fig. \ref{fig:burst_angle}. The hole location is relatively evenly distributed throughout the cap for small lifetimes (below 5~s or $0.3\times t_c$) and is concentrated towards the top of the cap (i.e. below 15 deg) for longer lifetimes. This concentration makes sense as the top of the cap is typically slightly thinner than the foot of the bubble. We can also notice from this data that the evaporation rate does not seem to play a role if defining the location where the hole is initiated. It is interesting to note that the transition towards the top of the bubble occurs at the same time as when evaporation starts to play a significant role (see Fig. \ref{fig:drain}).
These findings are similar to the ones obtained by \citet{champougny_life_2016} that observed a simultaneous increase of lifetime and decrease of $\theta$ (in our notation) as they increased surfactant concentration.

\end{document}